\documentclass[jcp,aps,final,preprint,groupedaddress,showpacs,floatfix,aip,reprint,amssymb,lengthcheck]{revtex4-1}
\usepackage{epsfig}
\usepackage[english]{babel}
\usepackage[fleqn]{amsmath}
\usepackage{subfig}
\usepackage{tabularx}
\usepackage{gensymb}
\usepackage{setspace}
\usepackage[usenames,dvipsnames,svgnames,table]{xcolor}

\begin{document}
\begin{spacing}{2}
\title{Theoretical study of line and boundary tension in adsorbed colloid-polymer mixtures}

\author{Jesper Koning}
\affiliation{Instituut voor Theoretische Fysica, Celestijnenlaan 200D, KU Leuven, B-3001 Leuven, Belgium}

\author{Yves Vandecan}
\affiliation{Instituut voor Theoretische Fysica, Celestijnenlaan 200D, KU Leuven, B-3001 Leuven, Belgium}

\author{Joseph Indekeu }
\affiliation{Instituut voor Theoretische Fysica, Celestijnenlaan 200D, KU Leuven, B-3001 Leuven, Belgium}

\date{\today}

\begin{abstract}
An extended theoretical study of interface potentials in adsorbed colloid-polymer mixtures is performed. To describe the colloid-polymer mixture near a hard wall, a simple Cahn-Nakanishi-Fisher free-energy functional is used. The bulk phase behaviour and the substrate-adsorbate interaction are modelled by the free-volume theory for ideal polymers with polymer-to-colloid size ratios $q=0.6$ and $q=1$. The interface potentials are constructed with help from a Fisher-Jin crossing constraint. By manipulating the crossing density, a complete interface potential can be obtained from natural, single-crossing, profiles. The line tension in the partial wetting regime and the boundary tension along prewetting are computed from the interface potentials. The line tensions are of either sign, and descending with increasing contact angle. The line tension takes a positive value of $10^{-14}$ - $10^{-12} N$ near a first-order wetting transition, passes through zero and decreases to minus $10^{-14}$ - $10^{-12} N$ away from the first-order transition. The calculations of the boundary tension along prewetting yield values increasing from zero at the prewetting critical point up to the value of the line tension at first-order wetting.

\end{abstract}

\maketitle

\section{Introduction}
A theoretical wetting study is presented with calculations of the line tension and the boundary tension of a model colloid-polymer mixture. The wetting properties of demixed colloid-polymer mixtures adsorbed at a glass wall have been investigated experimentally and theoretically, following the development of the confocal scanning laser microscopy technique for this system \cite{CSLM1,CSLM2,CSLM3}. This paper is an extension of a previous report in which only the line tension at first-order wetting was investigated \cite{YvesIndekeu}. Now, the line tension well into the partial wetting regime and the boundary tension along prewetting are considered for the same colloid-polymer model. 

The line tension is the excess free energy per unit length attributed to the contact line where three immiscible phases meet \cite{Indekeu,Indekeu2}. Here, the three phases are colloidal gas (a phase rich in polymers and poor in colloids), colloidal liquid (rich in colloids and poor in polymers), and a solid (substrate, which is a spectator phase). The solid phase is in mechanical equilibrium with the adsorbates and Young's law is applicable at the three-phase contact line \cite{Bonn}. Experimentally, a wetting transition of what seems to be first-order has been observed for a colloid-polymer mixture with a size ratio $q\approx 1$ \cite{Wijting,Wijting1}. However, to make a definitive statement about the order of this wetting transition more evidence is needed. Also, for size ratio $q\approx 0.6$ only complete wetting states have been observed \cite{Aarts,Aarts1}.

There is a possible extension of the first-order wetting transition off-coexistence where distinct surface phases are present \cite{Indekeu,Indekeu2,Perkovic,Boundarytension}. Here, only the colloidal gas is a stable bulk phase. However, microscopic and mesoscopic films of a colloidal liquid-like phase can be present at the wall. Coexistence of the two thin-film states is possible at a prewetting transition. A line of prewetting transitions starts at the wetting transition and ends in a prewetting critical point. Along this prewetting line, a boundary tension can be defined, which is conceptually related to the line tension. Two possible surface phases share a one-dimensional boundary, and the excess free energy per unit length attributed to this boundary is the boundary tension. In the prewetting critical point the boundary tension vanishes, whereas it takes the same value as the line tension in the wetting transition. 

This paper is set up as follows. First, the Cahn-Landau mean-field functional for colloid-polymer mixtures ~\cite{YvesIndekeu,YvesPhD,Inttensionwetting} is reviewed briefly and the interaction potentials and parameters are explained. Then, the interface potentials for the system at first-order wetting, in the partial wetting regime and along prewetting are presented and attention is given to the choice of the Fisher-Jin crossing constraint. The interface potentials are used to calculate the line and boundary tensions. Finally the results are summarised.

\section{A colloid-polymer mixture near a hard wall}
The colloid-polymer mixture near a hard wall can be described by the surface free energy functional
\begin{eqnarray}\label{CPMnearwall}
\gamma[\rho] = \int_{0}^{\infty} \mathrm{d} z \, \Bigl( f(\rho) -
\mu_c \rho(z) + p_c + m(\rho) \bigl( \frac{\mbox{d} \rho}{\mbox{d}
z} \bigr)^2 \Bigr) \nonumber \\ - h_1\rho_1 - \frac{1}{2}g
{\rho_1}^2.  \hspace{4cm}
\end{eqnarray}
The order parameter, $\rho(z)$, is the mean-field colloid number density at perpendicular distance $z$ from the wall, which is located at $z=-\sigma_c/2$, with $\sigma_c$ the colloid particle diameter. This means that the closest position the colloid can have is $z=0$. The colloid density is related to the dimensionless volume fraction through $\phi_c=(\pi/6)\sigma_c^3\rho$. The free energy functional $\gamma[\rho]$ is expressed in the Cahn-Nakanishi-Fisher form and consist of three parts. The first part is composed of the first three terms of the integrand, $f(\rho)-\mu_c\rho(z)+p_c$, which represent the negative of the excess pressure. In this relation, $\mu_c$ and $p_c$ are the equilibrium chemical potential and pressure at two-phase coexistence, respectively. The colloid-polymer mixture is treated as colloidal hard spheres and penetrable polymer spheres. Therefore, the free volume theory with ideal polymers without curvature effects can be applied. The free energy can be written as ~\cite{Phasstruct1}
\begin{equation}\label{Freevolfundamental}
F(N_c,V,T,z_p) = F_0(N_c,V,T) - p^R \alpha(\phi_c) V.
\end{equation}
Here $F_0(N_c,V,T)$ is the hard-sphere free energy, $p^R$ is the polymer reservoir pressure of ideal polymers -- which can be expressed in terms of the reservoir polymer volume fraction $\phi^r_p$ by $p^R=(6/\pi)\sigma_p^{-3}\phi_p^rk_BT$, with $\sigma_p$ the polymer ``diameter" related to the radius of gyration $R_g=\sigma_p/2$ -- and $\alpha(\phi_c)$ is the free-volume fraction available to polymers, which is a function of the colloid volume fraction, $\phi_c$. For $\alpha(\phi_c)$ an approximate expression is found in scaled particle theory ~\cite{Lekkerkerker,Lebowitz}. The hard-sphere free energy is given by the well known Carnahan-Starling expression \cite{CarnahanStarling}.

The second part of the free energy functional is the squared gradient term in the integrand, $m(\rho)(\frac{d\rho}{dz})^2$. The squared gradient is the leading term in the expansion of inhomogeneities of the order parameter \cite{Evansliqvap}. The coefficient of the squared gradient is given by $  m(\rho)\beta = \frac{\pi}{3} \int_{0}^{\infty} dr \,r^4 c(r,\rho) $ where $\beta=\dfrac{1}{k_BT}$ and $ c(r,\rho) $ represents the direct Ornstein-Zernike correlation function with colloid center-to-center distance $r$ and density $\rho$. The direct
correlation function is approximated by \cite{Theoryliquids,BradEvans}
\begin{equation}
c(r, \rho ) = \left\{
    \begin{array}{ll}
        0, & r \leq \sigma \\
        -\beta u(r), & r > \sigma,
    \end{array}
\right. \hspace{1.9cm}
\end{equation}
where $u(r)$ is an attractive interaction potential and arises from the overlap of the depletion zones around the colloid where the polymer is excluded. The two-colloid-particle interaction potential is given by the Asakura-Oosawa depletion potential \cite{AOAO,AOVrij}.

The terms outside the integrand, $-h_1\rho_1-\frac{1}{2}g\rho_1^2$, form the ``wall'' part of the surface free energy functional. These terms correspond to the contact interaction of the colloid-polymer mixture with the hard wall. Here, $h_1$ is the surface excess chemical potential, $g$ is the surface enhancement parameter, and $\rho_1=\rho(0)$, the colloid density at the wall. In the framework of the free-volume theory, $h_1=-\int_{0}^{\infty}dz\;U_2(z)$, where $U_2$ $=p^RV_2$, with  $V_2$ the volume overlap of the depletion zones of a colloidal particle and the hard wall \cite{Inttensionwetting,Aarts}. Also, $g=\int_{0}^{\infty}dz\int{d\vec{r}\;U_3(z,\vec{r})}$, where $U_3$ $=p^RV_3$, with $V_3$ the triple overlap of the excluded volumes of two colloids and the hard wall \cite{strictly}. Recently, Blokhuis and Kuipers argued that this representation of the surface enhancement parameter is incomplete \cite{Blokhuis}. They showed that an extra term has to be included, related to the missing colloid-colloid interactions near the wall. This has a significant impact on the physics of the colloid-polymer mixture. Blokhuis and Kuipers showed that inclusion of this term in the surface enhancement parameter causes wetting reversal at certain values for the aspect ratio $q=\dfrac{\sigma_p}{\sigma_c}$. This means that a drying transition would be observed instead of a wetting transition. Also, the nature of the wetting transition changes as it becomes a second-order phase transition, which leads to the absence of a prewetting regime and consequently no boundary tension can be defined. Another consequence of a second-order wetting transition would be that the line tension at the transition is zero. Although the inclusion of the extra term may constitute a more complete representation of the colloid-polymer mixture, the predictions exhibit poor agreement with experiments \cite{Blokhuis}. For this reason, and in view of our goal to complete the previous account \cite{YvesIndekeu} in which no consideration was made of this extra term in the surface enhancement parameter, this refinement to the theory is ignored. This means that a first-order wetting transition is considered here with a prewetting regime. Thus, for this study the free volume theory with ideal polymers as presented by Aarts et al. \cite{Inttensionwetting} is adopted. A phase diagram of this colloid-polymer system, including a prewetting line, was already presented there \cite{Inttensionwetting}. A thorough generic phase diagram for first-order wetting with representations of all the possible surface phases has been presented by Perkovi\`{c} et al. \cite{Boundarytension}.

\section{Interface Potentials}

\begin{figure}[h]
  \centering
  \includegraphics[width=0.45\textwidth]{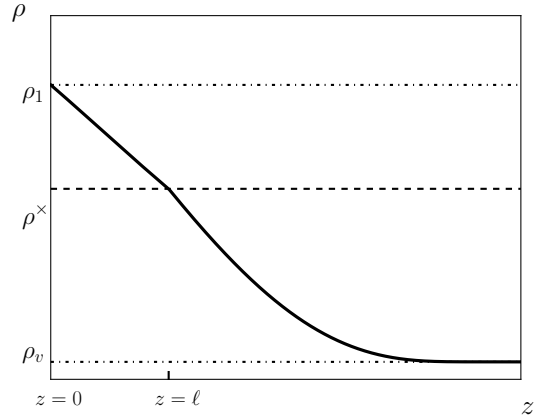}
  \caption{An example of a density profile on which the crossing constraint has been imposed. The profile starts with density $\rho_1$ at the wall and is constrained up to the crossing density $\rho^{\times}$. From there the constant of motion is set to zero and the density profile is the unconstrained one, ultimately reaching the bulk vapour phase density $\rho_v$. Note the discontinuity of the first derivative of $\rho(z)$ at $z=\ell$.}
  \label{fig:crossing}
\end{figure}

The Fisher-Jin interface potential \cite{FisherJin} is the excess free energy per unit area for a homogeneous liquid film of fixed thickness $\ell$ adsorbed at the spectator phase. Therefore, it can be regarded as a constrained, non-equilibrium surface free energy. The interface potential can be obtained from computing the surface free energy $\gamma[\rho]$ for the constrained density profile, the constraint being a crossing density $\rho^{\times}$ at a distance $\ell$ from the surface. An example of a density profile obeying the Fisher-Jin crossing constraint is shown in Fig. \ref{fig:crossing}.
In the previous report \cite{YvesIndekeu}, it was argued that single-crossing profiles are insufficient to map the complete interface potential. For single-crossing profiles, the density follows the constrained profile up to the value of the crossing density with the corresponding constant of motion. After the crossing density is reached the density profile is the unconstrained profile, for which the constant of motion is zero. The double-crossing profiles follow the unconstrained equilibrium phase portrait only after the crossing density is encountered for the second time. For a graphical explanation of the single-crossing and double-crossing profiles, see Fig. 1b in Ref. \cite{YvesIndekeu}.  

Here, we argue that the double-crossing profiles can be avoided if the crossing density is chosen more judiciously. In this section the interface potentials for partial wetting and for prewetting are considered.
\subsection{Partial Wetting}

The interface potential of a colloid-polymer mixture with $q=1$ near first-order wetting with $\phi_c^{\times}=0.03$ -- where the crossing volume fraction is related to the crossing density through $\phi_c^{\times}=(\pi/6)\sigma_c^3\rho^{\times}$ -- was already reported in Ref \cite{YvesIndekeu,YvesPhD}. The necessary equations are repeated briefly. For the optimal density profile the Euler-Lagrange equation has to be solved, while complying with the boundary conditions and crossing constraints. This means that after the density profile reaches the crossing density the constant of motion is zero, whereas leading up to the crossing density the constant of motion can be either positive or negative.
The Euler-Lagrange equation reads
\begin{equation}
2m\dfrac{d^2 \rho}{dz^2}=\dfrac{df(\rho)}{d\rho}-\mu_c
\end{equation}
and the following boundary condition applies
\begin{equation}
-h_1-g\rho_1=2m\dfrac{d\rho}{dz}(0).
\end{equation}
The constant of motion is
\begin{equation}
m(d\rho/dz)^2-f(\rho)+\mu_c\rho-p_c=E.
\end{equation}
\begin{figure}[h]
  \centering
    \includegraphics[width=0.45\textwidth]{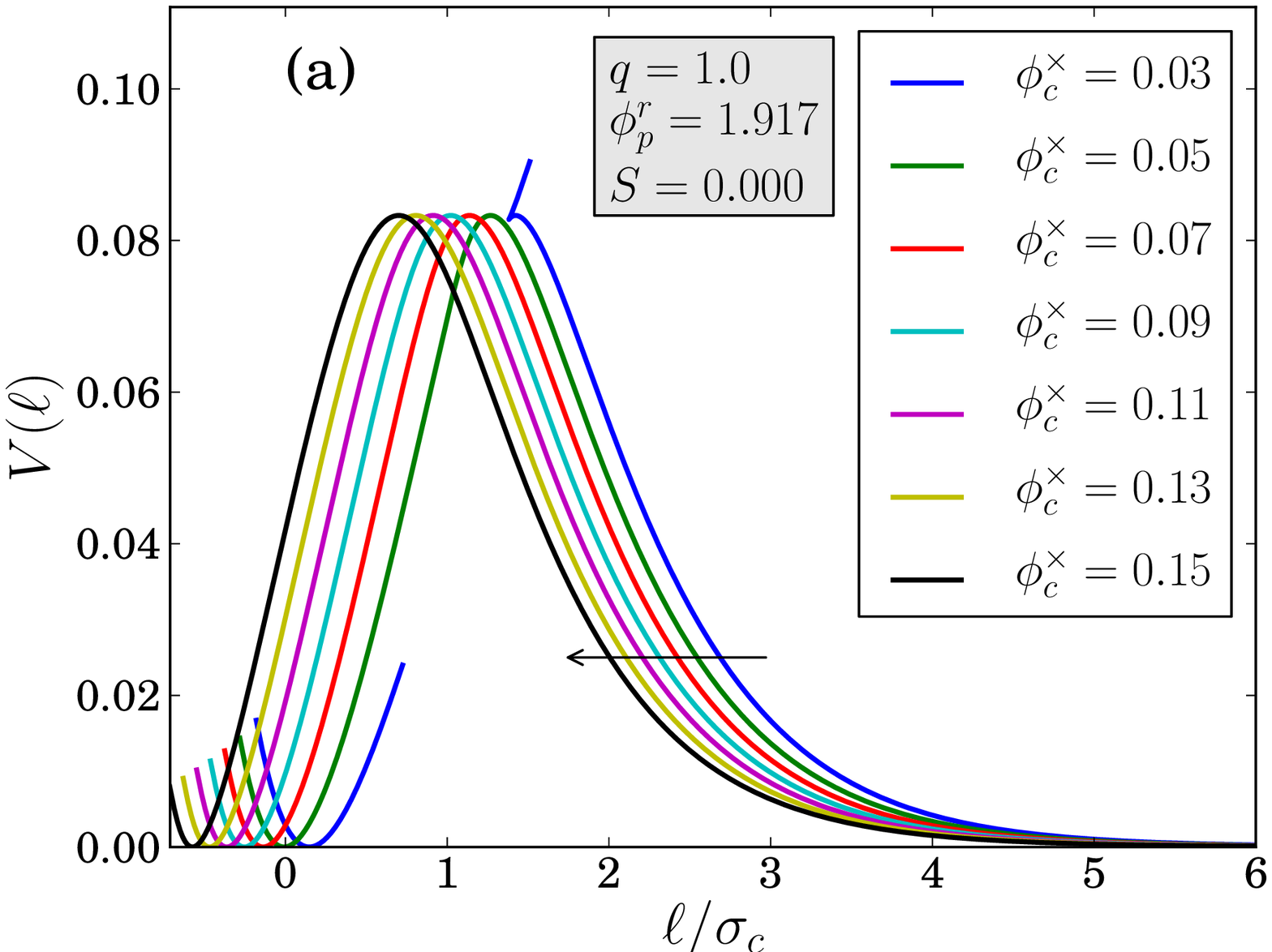}
    \includegraphics[width=0.45\textwidth]{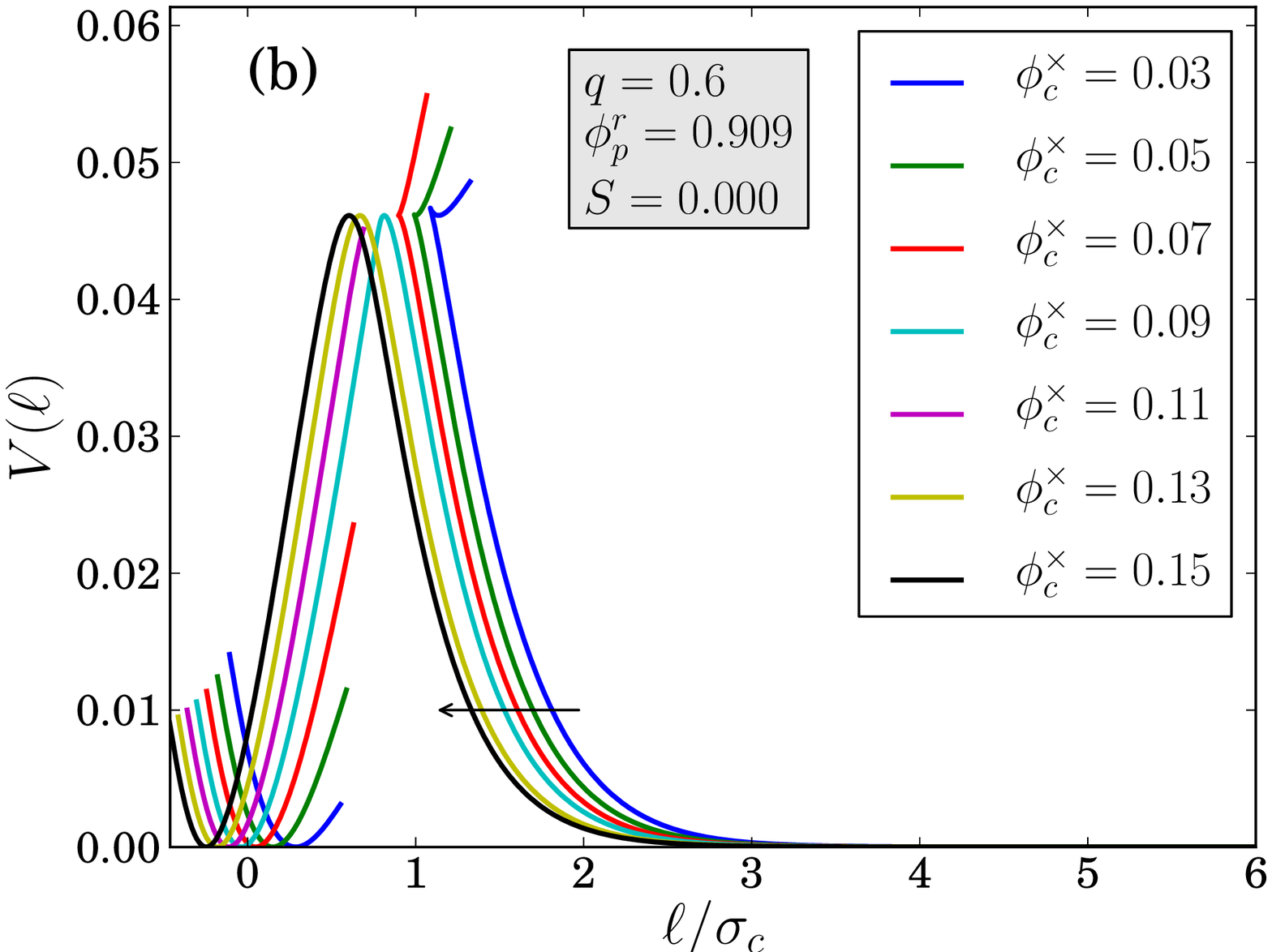}
  \caption{Interface potentials, in units of $\frac{k_BT}{(\pi \sigma_c/6)^2}$, near the wetting transition ($S=0$), which can be obtained via a graphical construction of the phase portrait \cite{DeGennes,Cahn}. (a) For aspect ratio $q=1$, the interface potentials are incomplete for crossing colloid volume fraction $\phi_c^{\times}=0.03$, where $\phi_c^{\times}=(\pi/6)\sigma_c^3\rho^{\times}$. The interface potentials are complete for $\phi_c^{\times}=0.05$ and higher. (b) For aspect ratio $q=0.6$, the interface potentials become complete for crossing volume fraction $\phi_c^{\times}=0.09$ and higher. The arrows indicate the evolution of the control parameter $\phi_c^{\times}$ from low to high.}
\label{fig:FOwet}
\end{figure}
\\
Double-crossing profiles were necessary for $\phi_c^{\times}=0.03$ and $q=1$, as the interface potential was incomplete \cite{incomplete}. In Fig. \ref{fig:FOwet} the interface potentials of single-crossing density profiles at the wetting transition, where the so-called spreading coefficient $S\equiv \gamma_{SG}-(\gamma_{SL}+\gamma_{LG})$ is zero, are plotted. It can be seen in Fig. \ref{fig:FOwet}(a) that raising $\phi_c^{\times}$ to $0.05$ takes care of the incomplete interface potential and allows one to obtain a fully smooth and complete one. Further increasing the crossing density causes a shift of the interface potential to lower values of $\ell$. However, the shape of the interface potentials remains the same. The local minimum associated with the thin film starts out at a positive $\ell$ for low values of the crossing density. As the crossing density increases the location of the minimum shifts towards lower values of $\ell$, even reaching negative values. This is not problematic, since the collective coordinate $\ell$ is not the physical thickness of the wetting layer, but a mathematical location which may lie behind the wall \cite{BonnJosPos}. In such cases, in the strictest sense, there is no wetting film, but rather just some modest adsorption at the wall. For first-order wetting and $q=0.6$, which is depicted in Fig. \ref{fig:FOwet}(b), the interface potentials become completely smooth functions for $\phi_c^{\times} \approx 0.09$. High crossing densities lead to familiar interface potentials \cite{BonnJosPos,Indekeu}, with an ordinary barrier without a discontinuity at the maximum between the thin-film minimum and the complete-wetting layer. For low crossing densities, a second branch near the local maximum (at $\ell/\sigma_c \approx 1$) is visible. These branches should be ignored because only the minimal value of $V$ is of interest. The incomplete interface potentials are disregarded for the calculation of the line tension. In this respect, our present approach is a refinement of that taken in \cite{YvesIndekeu}.\\

\begin{figure}[h]
\centering
\includegraphics[width=0.45\textwidth]{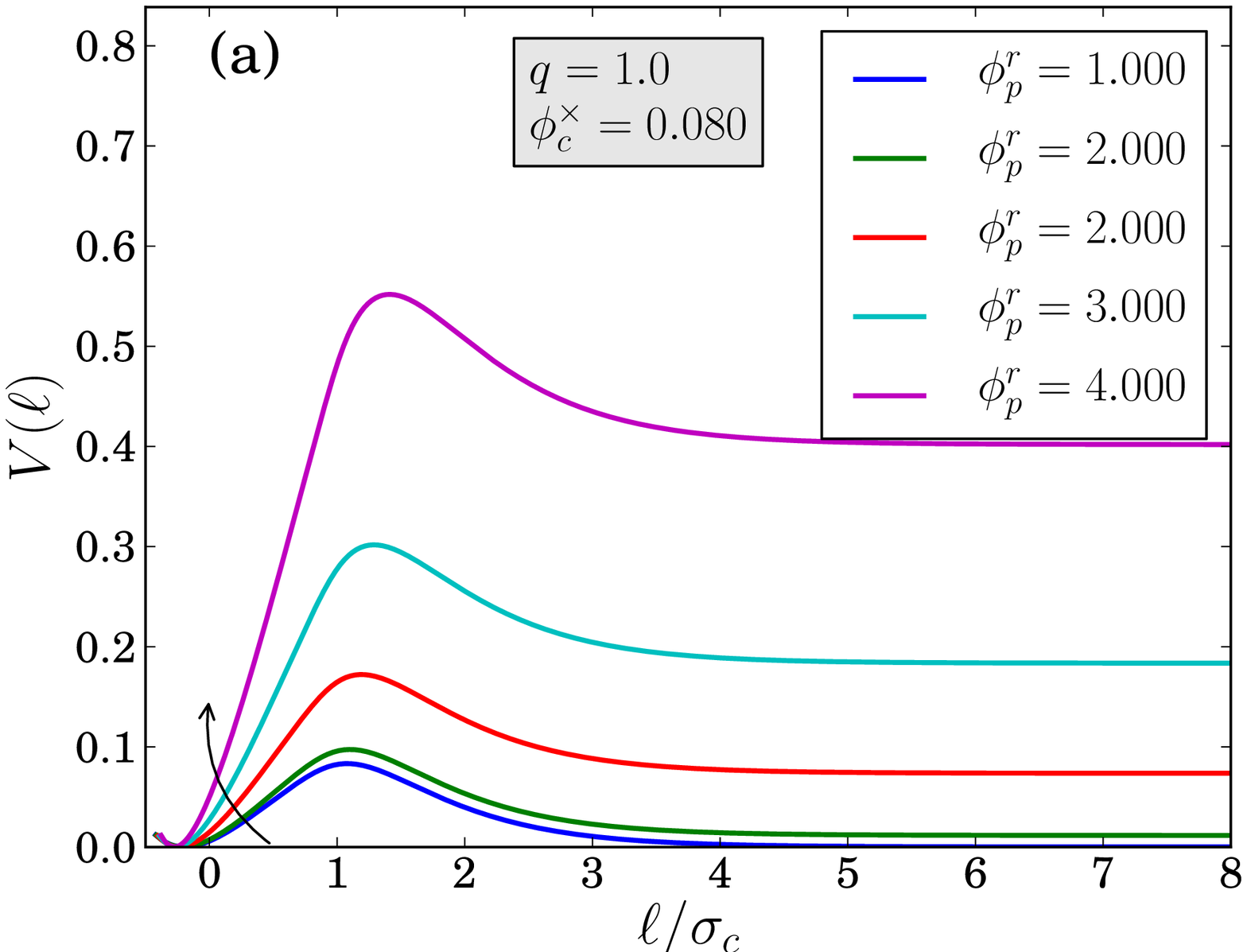}
\includegraphics[width=0.45\textwidth]{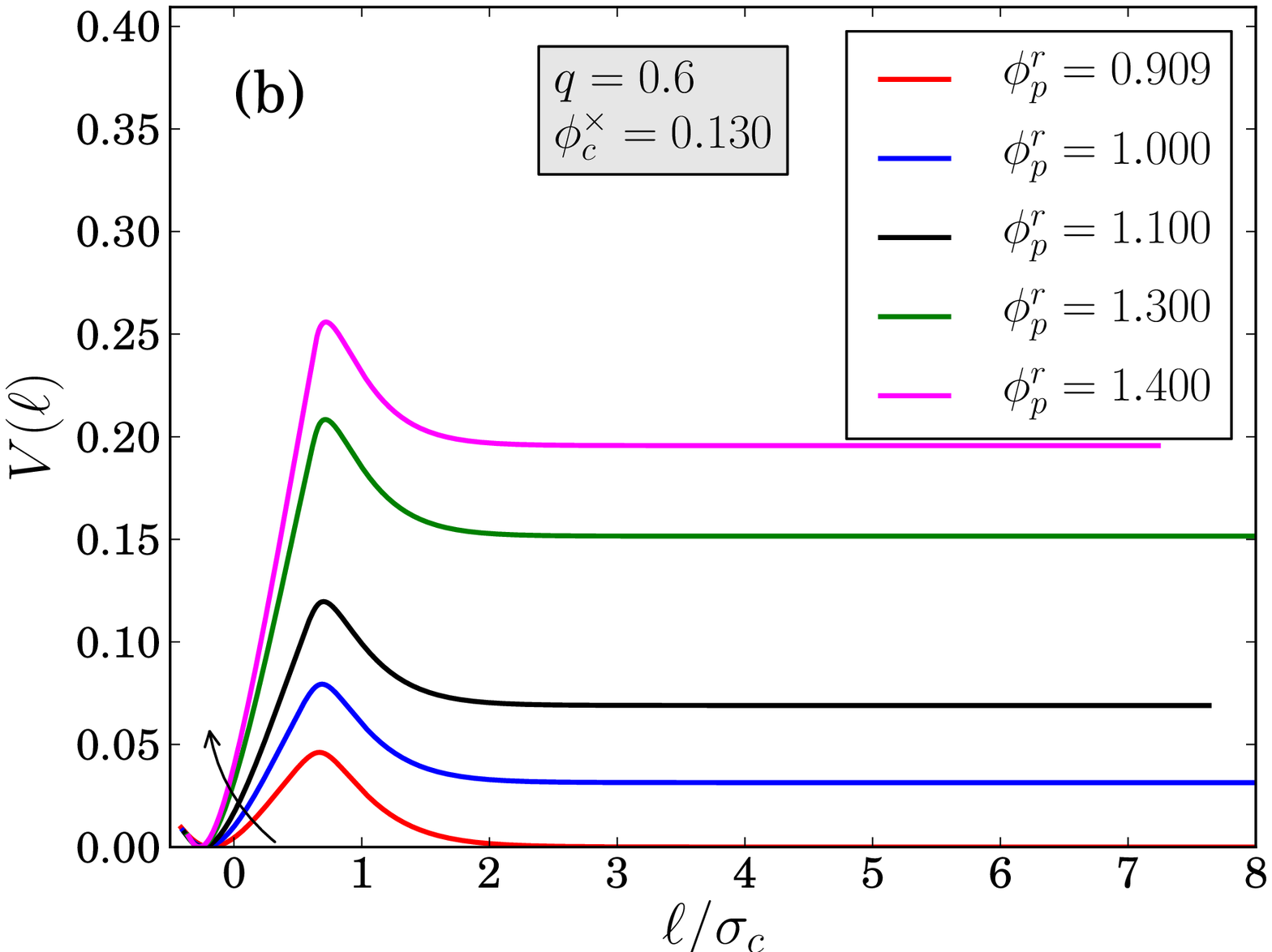}
\caption{Overview of interface potentials, in units of $\frac{k_BT}{(\pi \sigma_c/6)^2}$, in the partial wetting regime for (a) $q=1$ at $\phi_c^{\times}=0.08$  and (b) $q=0.6$ at $\phi_c^{\times}=0.13$. The arrows indicate the evolution of the control parameter $\phi_p^r$ from low to high.}
\label{fig:InterfaceWetting}
\end{figure}

The interface potentials for states in the partial wetting regime are plotted in Fig. \ref{fig:InterfaceWetting}. For the Fisher-Jin crossing volume fraction values $\phi_c^{\times} = 0.13$ and $0.08$ for $q=0.6$ and $q=1.0$, respectively, the following fitting curve is proposed as an instructive analytical representation of the interface potential.

\begin{eqnarray}\label{VlfitH}
V(\ell) =  \left\{
    \begin{array}{lll}
        a_2(\ell - \ell_1)^2 + a_3(\ell - \ell_1 )^3 + \\
        a_4(\ell-\ell_1)^4, &   0 < \ell < \ell_a \\ \\
        b_1 e^{-\ell/\xi} + b_2 e^{-2\ell/\xi}
         + b_0, & \ell_a < \ell
    \end{array}
\right. \nonumber \\
\end{eqnarray}
This fit consists of two parts. The first part is a Taylor expansion about the thin-film minimum at $\ell_1$ up to third or fourth order. The second part reflects the exponential tail converging to minus the proper spreading coefficient, $-S=b_0$. This tail is approximated by $e^{-\ell/\xi}, e^{-2\ell/\xi}, \cdots$ with $\xi$ the bulk correlation length. This fit introduces an artificial singularity at $\ell=\ell_a$, whereas the interface potentials are smooth throughout. Table \ref{Interfacepotfit1} reports on some typical fitting parameter values near and far from the wetting transition.  

\begin{table}[h]
\begin{center}
\begin{tabular}{|c|c|c|c|c|c|c|c|}
\hline
\textbf{$\phi_p^r$} & \textbf{$q$} & \textbf{$\ell_{1}/\sigma_c$} & \textbf{$a_2$}   & \textbf{$a_3$} & \textbf{$a_4$} & \textbf{$\ell_{a}/\sigma_c$} \\
\hline
1.917 &  1    & -0.19718  & 0.15357 & -0.08267 & 0.00191 & 1.0044 \\
\hline
3.0  &  1  &  -0.2611   & 0.40599 & -0.19963 & 0.01228 & 1.0603 \\
\hline
\end{tabular}
\end{center}

\begin{center}
\begin{tabular}{|c|c|c|c|c|c|c|c|c|c|c|c|}
\hline
\textbf{$\phi_p^r$} & \textbf{$q$} &  \textbf{$b_0$}   &   \textbf{$b_1$}   & \textbf{$b_2$}    & \textbf{$\xi/\sigma_c$}    \\
\hline
1.917 &  1    &  0.000041 & 0.77398  & -1.794  & 0.70779  \\
\hline
3.0  &  1  &    0.18356   &  1.40738  & -4.18225   & 0.7226 \\
\hline
\end{tabular}
\end{center}
\caption{Illustrative $V(\ell)$-fitting parameters in the partial wetting regime for $q=1$ ($\phi_c^{\times} = 0.08$).}\label{Interfacepotfit1}
\end{table}

\subsection{Prewetting towards complete wetting}

\begin{figure}[h]
\centering
\includegraphics[width=0.45\textwidth]{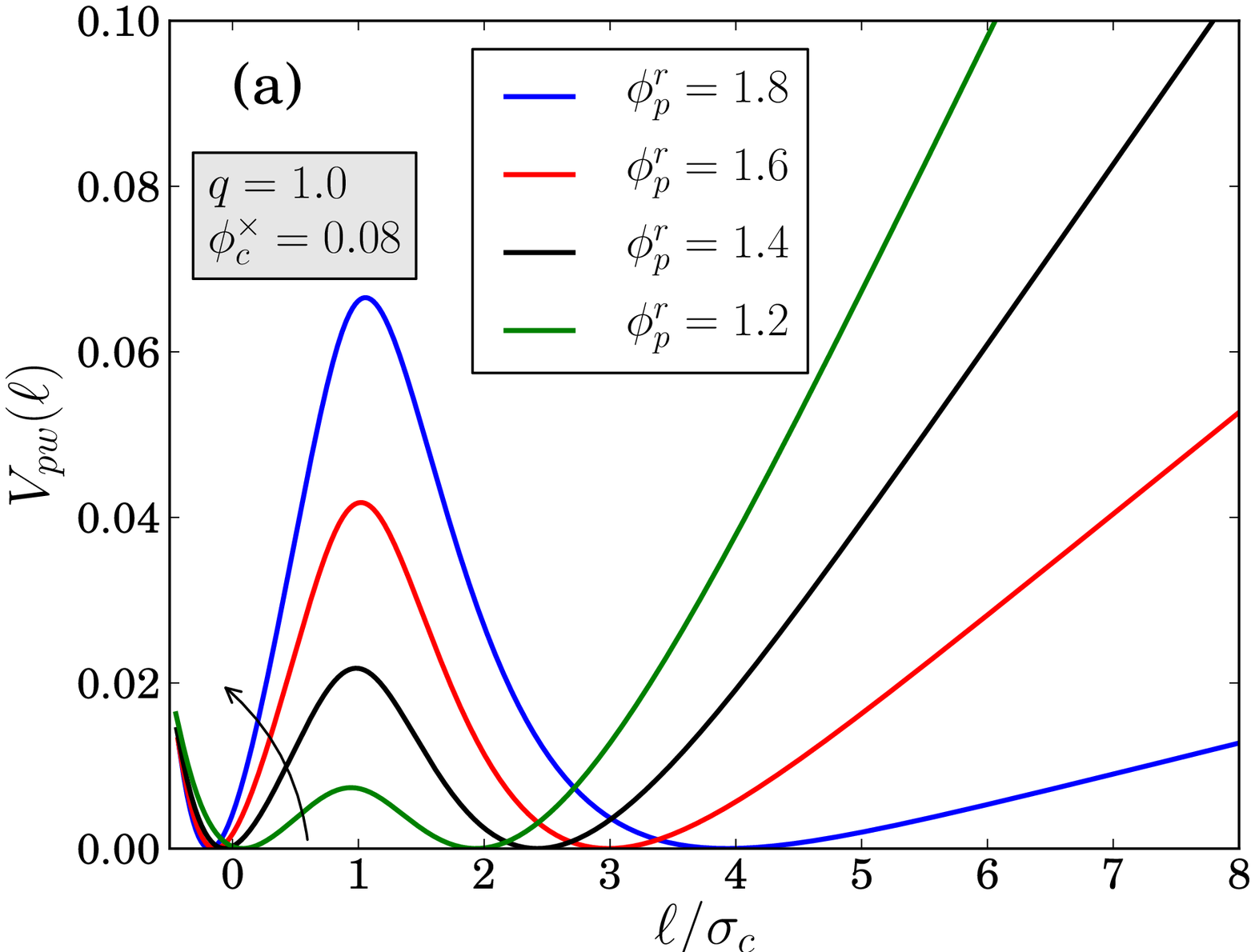}
\includegraphics[width=0.45\textwidth]{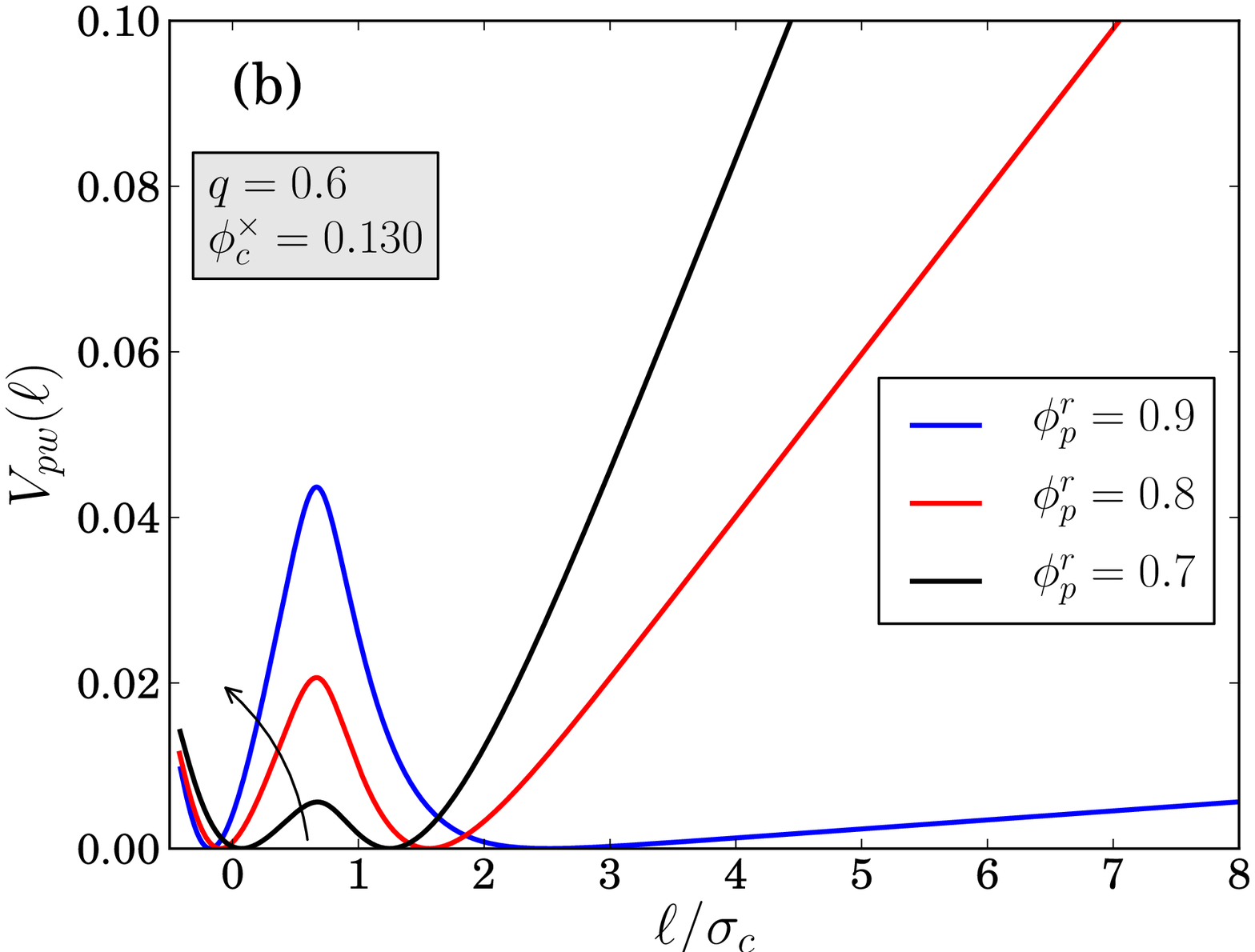}
\caption{Overview of interface potentials, in units of $\frac{k_BT}{(\pi \sigma_c/6)^2}$, along prewetting for (a) $q=1$ at $\phi_c^{\times}=0.08$  and (b) $q=0.6$ at $\phi_c^{\times}=0.13$. The arrows indicate the evolution of the control parameter $\phi_p^r$ from low to high.}
\label{fig:PreWet}
\end{figure}

The interface potential that describes the prewetting transition takes the form
\begin{equation}
V_{pw}(\ell) = V(\ell) + \gamma_{LG}H \ell/\xi,
\end{equation}
where $V(\ell)$ is, with minor modifications, the interface potential at the wetting transition and $H$ is a dimensionless field which takes the system off of two-phase equilibrium and $\gamma_{LG}$ is the liquid-gas surface tension at zero field. The factor $\gamma_{LG}H/\xi$ takes the value of the constant of motion at large $\ell$. Along the prewetting line as $H \rightarrow 0$ the thick-film thickness increases to infinity, and the interface potential converges to the first-order wetting interface potential $V(\ell)$. As can be seen in Fig. \ref{fig:PreWet} the interface potential in the prewetting regime displays two minima, congruent with two surface phases. The first minimum corresponds to the thin film and the second corresponds to the thick film. Table \ref{Interfacepotfit2} reports typical fitting parameter values for several state parameters along prewetting. Note that the Taylor expansion around the minimum is truncated after third order.\\ 
\\

\begin{table}[h]
\begin{center}
\begin{tabular}{|c|c|c|c|c|c|c|}
\hline
\textbf{$\phi_p^r$} & \textbf{$q$} & \textbf{$\ell_{1}/\sigma_c$} & \textbf{$a_2$}   & \textbf{$a_3$} & \textbf{$\ell_{a}/\sigma_c$} \\
\hline
1.2 &   1   & 0.081   & 0.0319 & -0.02567 &  0.9087 \\
\hline
1.91 &  1    & -0.19623  & 0.15034 & -0.0782 &   1.1538\\
\hline
\end{tabular}
\end{center}

\begin{center}
\begin{tabular}{|c|c|c|c|c|c|c|c|c|c|c|c|c|}
\hline
\textbf{$\phi_p^r$} & \textbf{$q$} &  \textbf{$b_0$}   &   \textbf{$b_1$}   & \textbf{$b_2$}    & \textbf{$\xi/\sigma_c$} & \textbf{$\gamma_{LG}H/\xi$}   \\
\hline
1.2 &  1 &  -0.08969   &  0.35502  & -0.44494 & 0.81564 & 0.03125 \\
\hline
1.91 & 1  &   -0.001  &  0.78637   & -1.86227  & 0.70257 & 0.0001429 \\
\hline
\end{tabular}
\end{center}
\caption{Illustrative $V_{pw}(\ell)$-fitting parameters along the prewetting line for $q=1$ ($\phi_c^{\times} = 0.08$). $\gamma_{LG}H/\xi$ is given in terms of $\frac{k_BT}{(\pi/6)^2\sigma_c^3}$.}\label{Interfacepotfit2}
\end{table}

\section{Line and boundary tension}

It was previously explained that the excess free energy per unit length of the surface inhomogeneity can be treated within the interface displacement theory \cite{YvesIndekeu}. The line tension is a functional of the displacement of the interface perpendicular to the substrate, and minimising this functional yields the equilibrium line tension \cite{Indekeu1,Indekeu2}. In the squared gradient approximation, this leads to the following expression for the line tension \cite{Indekeu2}
\begin{equation}
\tau=(2\gamma_{LG})^{1/2}\xi \int_{\ell_1}^{\infty}\mathrm{d}\tilde{\ell}\Big[V(\tilde{\ell})^{1/2}-(-S)^{1/2}\Big],
\label{eq:linetension}
\end{equation}
where $\gamma_{LG}$ is the liquid-gas surface tension, $\tilde{\ell}=\ell/\xi$ with $\xi$ the correlation length, and $-S$ is minus the spreading coefficient, which takes the value of $\lim_{\ell\rightarrow \infty}V(\ell)$. The lower integration limit, $\ell_1$, is the thin-film minimum.
\begin{figure}[h]
\centering
\includegraphics[width=0.45\textwidth]{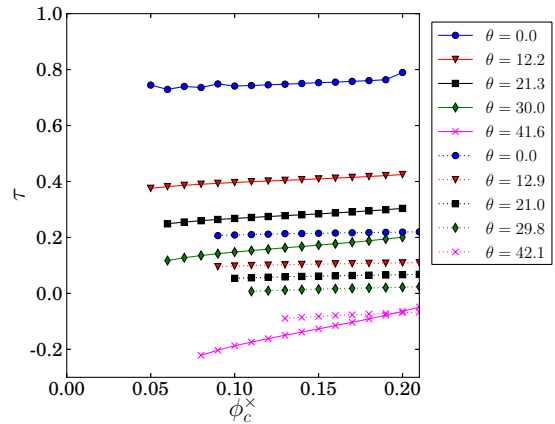}
\caption{Line tension, in units of $\frac{k_BT}{(\pi /6)^2\sigma_c}$, as a function of the crossing volume fraction for contact angles $\theta$ (degrees) up to about $40\degree$ for both $q=1$ (full line) and $q=0.6$ (dotted line).}
\label{fig:Linetension}
\end{figure}

Via the spreading coefficient the contact angle can be calculated. In Fig. \ref{fig:Linetension} the line tension is plotted versus the Fisher-Jin crossing volume fraction for several values of the contact angle. The line tension is only weakly dependent on the chosen constraint, as was expected from the interface potentials seen in Fig. \ref{fig:FOwet}. It can be observed that for $q=1$ the line tension rises slightly with increasing values of the crossing density. Farther in the partial wetting regime, as the contact angle increases, the dependence of the line tension on the crossing density becomes more apparent. For $q=0.6$ variation of the line tension with the chosen crossing density is almost non-existent, as it should be in a reliable theoretical model, as it almost looks like a flat line regardless of the proximity to the wetting transition.

For typical colloidal diameters of $10-100 nm$, the computed line tensions are of the order $10^{-14}-10^{-12} N$ near first-order wetting, pass through zero and reach minus their values at first-order wetting for larger contact angles. The line tension values for $q=0.6$ are lower than those for $q=1$, especially near the wetting transition. This can be easily comprehended from the larger area under the $V(\ell)$ curve, as already pointed out near first-order wetting \cite{YvesPhD}. However, with increasing contact angle the difference in line tension decreases and for $\theta\approx 40\degree$ the line tensions are equal. It can be observed in Fig. \ref{fig:Linetension} that the crossing density necessary to provide a complete interface potential, and thus a value for the line tension, increases with increasing contact angle.

\begin{figure}[h]
\centering
\includegraphics[width=0.45\textwidth]{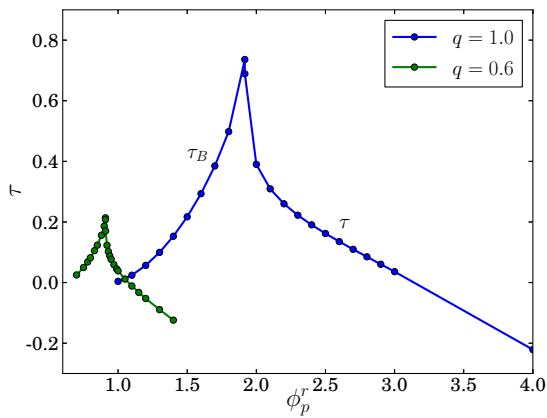}
\caption{Boundary ($\tau_B$) and line tension ($\tau$), in units of $\frac{k_BT}{(\pi /6)^2\sigma_c}$, for $q=1$ at $\phi_c^{\times}=0.08$ and $q=0.6$ at $\phi_c^{\times}=0.13$. The boundary tension in the prewetting critical point is zero. The line and boundary tensions are maximal and equal at the wetting transition.}
\label{fig:Boundarytension}
\end{figure}
Like for the line tension, an expression for the boundary tension can be found by using the interface displacement model. This leads to \cite{Indekeu2} 
\begin{equation}
\tau_B=(2\gamma_{LG})^{1/2}\xi\int_{\ell_1}^{\ell_2}\mathrm{d}\tilde{\ell}(V_{pw}(\tilde{\ell}))^{1/2}.
\label{eq:boundarytension}
\end{equation}
Here, $\ell_1$ and $\ell_2$ are the thin-film and thick-film minima, respectively. The boundary tension for $q=1$ and $\phi_c^{\times}=0.08$ as well as for $q=0.6$ and $\phi_c^{\times}=0.13$ is plotted against the reservoir volume fraction and the familiar lambda shape \cite{Indekeu,Perkovic,Boundarytension} is found. In the prewetting critical point, the thin and thick-film minimum come together and the boundary tension is zero. The boundary tension reaches the line tension at the wetting transition upon increasing the reservoir volume fraction. {red}The physical parameters at the critical point, the wetting transition, and the prewetting critical point are summarised in Table \ref{tab:summary}.
\begin{table}
\begin{center}
\begin{tabular}{|c|c|c|c|c|c|c|}
\hline
$q$ & $\phi_p^r$ (cp) & $\phi_c$ (cp) & $\phi_p^r$ (pwcp) & $\phi_c$ (pwcp) & $\phi_p^r$ (wet) & $\tau$\\
\hline
0.6 & 0.4879 & 0.1878 & 0.6007 & 0.0296  & 0.9086 & 0.2134 \\
\hline
1.0 & 0.6364 & 0.1040 & 0.9552 & 0.00681 & 1.917 & 0.7360 \\
\hline
\end{tabular}
\end{center}
\caption{For $q=1$ and $q=0.6$, the location of the critical point (cp), prewetting critical point (pwcp), and wetting transition (wet) are listed in terms of the polymer and colloid volume fractions $\phi_p^r$ and $\phi_c$. Also the value of the line tension $\tau$ at wetting is given, in units of $\frac{k_BT}{(\pi/6)^2\sigma_c}$. }
\label{tab:summary}
\end{table} 

\section{Conclusion}

For an adsorbed colloid-polymer mixture, a previous study of the interface potentials and the line tension at the wetting transition using a Cahn-Nakanishi-Fisher functional and an interface displacement model was extended to include the partial wetting and the prewetting regime. The colloid-polymer mixtures are modelled with the free-volume theory applied to ideal polymers without curvature effects. Previously, it was argued that double-crossing profiles were necessary to remedy incomplete interface potentials. Here, it was shown that, near the wetting transition, by increasing the crossing density up to $\phi_c^{\times}=0.05$ and $\phi_c^{\times}=0.09$ for $q=1$ and $q=0.6$, respectively, a smooth and complete interface potential can be obtained. Increasing the crossing density even further brings about a shift of the interface potential to lower $\ell$. Farther in the partial wetting regime, the minimal crossing density to obtain complete interface potentials increases.

At the wetting transition and $q=1$, the line tension increases only slightly with increasing crossing densities. This behaviour becomes more pronounced further into the partial wetting regime, where the line tension becomes negative. For $q=0.6$ the line tension curve is almost constant at the wetting transition and throughout the partial wetting regime. The computed line tensions have a value of $10^{-14}-10^{-12} N$ near first-order wetting, pass through zero and become negative for larger contact angles. For a fixed particle size, close to the wetting transition, the line tension for $q=0.6$ is significantly lower than for $q=1$, whereas they are equal at a contact angle of approximately $40 \degree$. The computed boundary tension along prewetting ranges from zero at the prewetting critical point to the line tension value at the wetting transition.
\end{spacing}

\section{Acknowledgements}
J.K and J.I are supported by KU Leuven Research Grant OT/11/063.

\newpage

\bibliography{CPM}

\begin{thebibliography}{32}
\bibitem [{inc()}]{incomplete}%
  \BibitemOpen
  \href@noop {} {}\bibinfo {note} {For several combinations of the initial
  density and the crossing density, the first derivative of the density
  $\frac{d\rho}{dz}$ becomes zero before the profile can reach the crossing
  density. Therefore, there is a gap in the interface potential, which is then
  called `incomplete'.}\BibitemShut {Stop}%
\end{thebibliography}
\bibliographystyle{aipnum4-1}

\end{document}